\documentclass{iaus}
\usepackage{wrapfig,graphicx}

\def \ls           {\hbox{L$_{\odot}$}}
\def \ms           {\hbox{M$_{\odot}$}}

\title[Star Formation Law] 
{The Global Star Formation Law: from Dense Cores to Extreme Starbursts}

\author[Y. Gao] 
{Yu Gao
}

\affiliation{Purple Mountain Observatory,
2 West Beijing Road, Nanjing 210008; and National Astronomy
Observatory, Chinese Academy of Sciences, Beijing, P.R. China 
 yugao@pmo.ac.cn,ygao@nrao.edu}

\pubyear{2007}
\volume{237}  
\pagerange{}
\date{?? and in revised form ??}
\jname{Triggered Star Formation in a Turbulent ISM}
\editors{Bruce G. Elmegreen \& J. Palous, eds.}
\begin{document}

\maketitle

\begin{abstract}

Active star formation (SF) 
is tightly related to the dense molecular gas in the giant 
molecular clouds' dense cores. Our HCN (measure of the dense molecular 
gas) survey in 65 galaxies (including 10 ultraluminous galaxies) 
reveals a tight linear correlation between HCN and IR (SF 
rate) luminosities, whereas the correlation between IR and CO (measure 
of the total molecular gas) luminosities is nonlinear. This suggests 
that the global SF rate depends more intimately upon the 
amount of dense molecular gas than the total molecular gas content. 
This linear relationship extends to both the 
dense cores  in the Galaxy and the hyperluminous extreme starbursts 
at high-redshift. Therefore, the 
global SF law in dense gas appears to be linear all the way 
from dense cores to extreme starbursts, spanning over nine orders 
of magnitude in IR luminosity.

\keywords{stars: formation -- ISM: molecules -- infrared: galaxies --
 radio lines: galaxies -- galaxies: high-redshift}
\end{abstract}

\section{Introduction}
 
Schmidt (1959) law of star formation (SF) was first formulated 
in terms of local SF rate $SFR$ proportional to the 
HI gas density with a power index n ($SFR \sim \rho^n$, n$\sim$1-3) 
as the atomic gas was then known as
the major component of interstellar gas reservoir to possibly
form stars. More than a decade later, 
observations of the CO line emission in the Milky Way and external galaxies
enabled by millimeter (mm) astronomy suggest that stars are forming in the
giant molecular clouds (GMCs). From 80's, e.g., Kennicutt (1983) 
found little evidence in parameterizing the global SF 
law in external galaxies in terms of the total HI gas. 
But he succeeded in terms of the disk-averaged surface $SFR$ and 
total surface gas densities
of both HI and H$_2$ (Kennicutt 1989). However, a well determined power 
index n was still not practical. Other researchers also obtained 
a wide range of the power index from 1 to 3. Recently, Kennicutt (1998) 
seemed to obtain a well determined slope of 1.4 though discrepancy 
exists (e.g., Heyer et al. 2004). 

We here show that the SF law in terms of the dense gas is 
a rather simple linear relation (Gao \& Solomon 2004a, GS04a) and 
is straight-forward to understand in what 
we have learned so far from the physics of SF. The cool atomic 
gas has the potential to convert into molecular form to possibly 
provide the fuel to make stars, yet even the bulk of molecular gas and
the most regions of GMCs are not making stars except for those tiny 
dense cores.  Great Observatories indeed begin to directly 
link/picture the heavily obscured dusty regions 
of active SF, embedded in GMCs as mapped by mm/sub-mm telescopes, 
to the current  massive SF.

\section{Correlations Among FIR, HCN \& CO In Galaxies}

 
\begin{wrapfigure}[]{r}{0.55\textwidth}
\vskip -0.2in
\includegraphics[height=2.8truein]{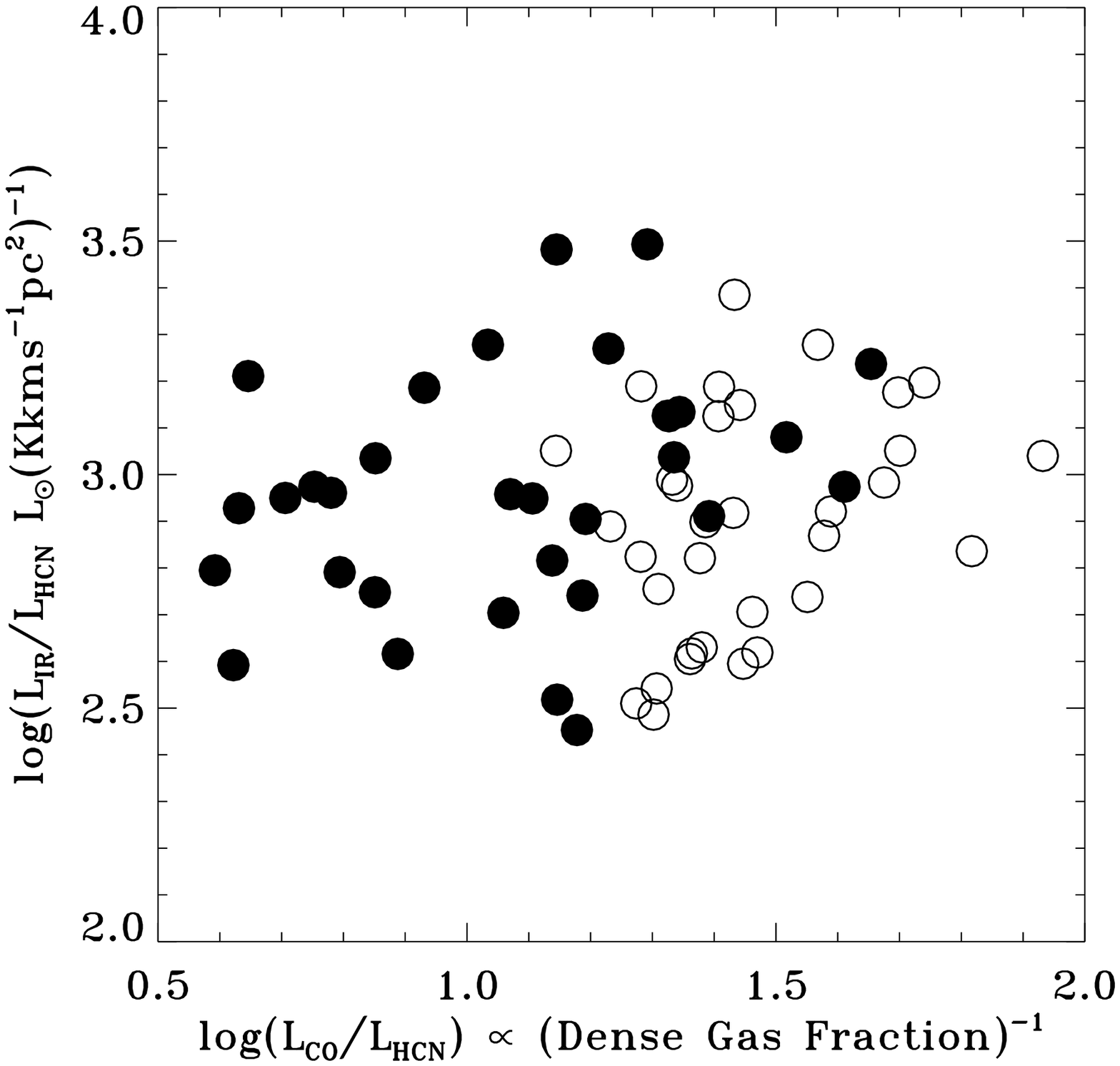}
\includegraphics[height=2.8truein]{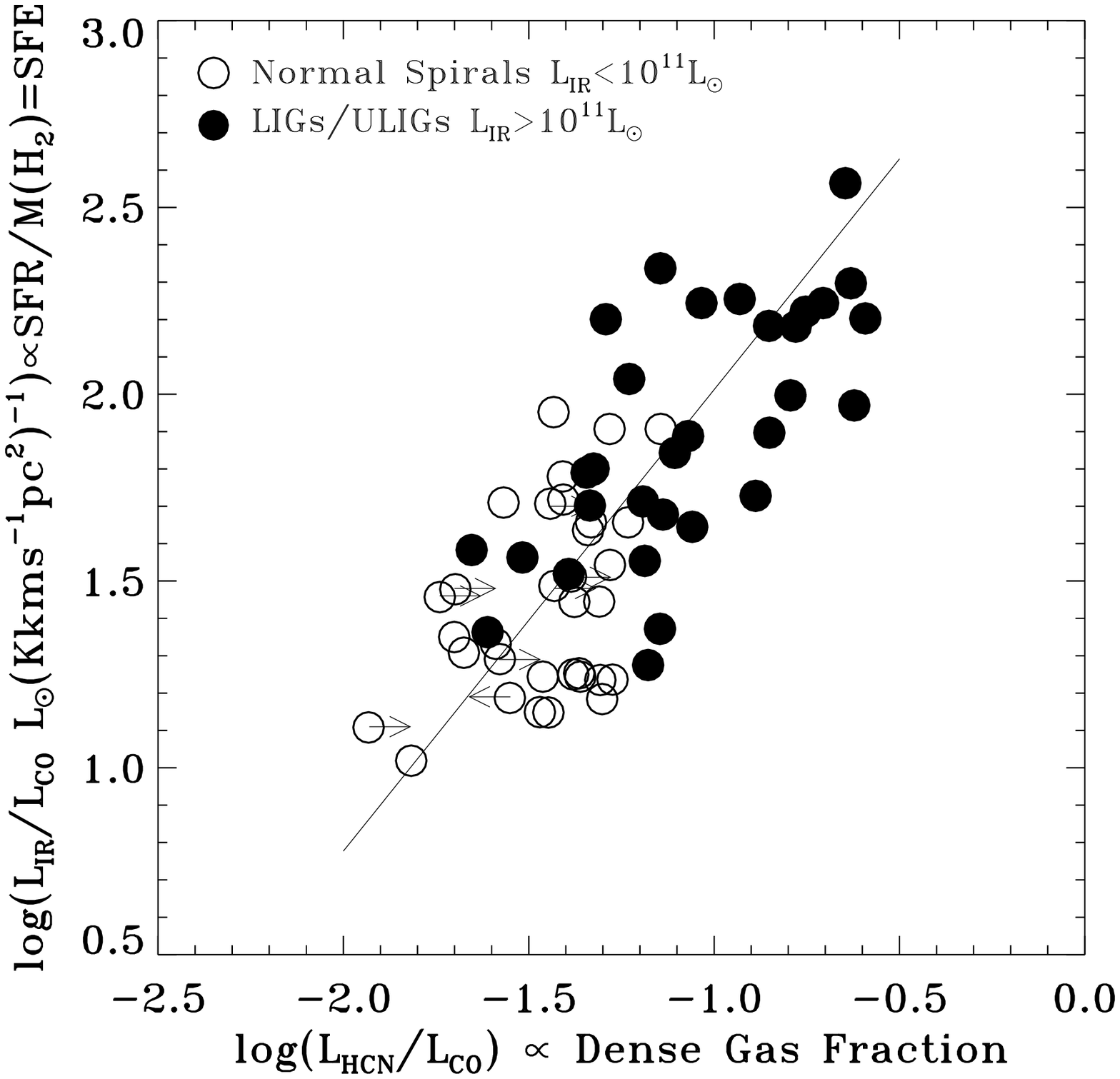}
\vskip -0.1in
\caption{ No correlation is observed 
between $L_{\rm IR}/L_{\rm HCN}$ and
$L_{\rm CO}/L_{\rm HCN}$ (top), yet there is still strong correlation between
$L_{\rm HCN}/L_{\rm CO}$ and $L_{\rm IR}/L_{\rm CO}$. 
The IR--CO correlation can simply be a result of
the much better correlations of IR--HCN and HCN--CO.}
\end{wrapfigure}

High-dipole moment molecules such as HCN trace more than order of
magnitude higher gas density than that of CO. A major HCN survey 
in a wide range of 65 galaxies, 
including 10 ultraluminous infrared galaxies (ULIRGs), tripled
the sample of galaxies with global HCN measurements 
(Gao \& Solomon 2004b). Analysis of the various relationships among 
the global HCN, CO, and FIR luminosities can be statistically 
conducted for the first time (GS04a).

The strong IR--HCN correlation is linear and extremely tight 
over 3 orders of magnitude in luminosity, when compared to the 
non-linear IR--CO correlation (GS04a). While the high  
luminosity of ULIRGs 
requires an  elevated SF efficiency of the total molecular gas indicated by 
$L_{\rm IR}/L_{\rm CO}$, the $SFR$ per unit of {\it dense} molecular 
gas, the SF efficiency of the {\it dense} molecular gas indicated 
by ($L_{\rm IR}/L_{\rm HCN}$) is  almost constant 
and independent of the IR luminosity or total $SFR$.
Further, GS04a find the surprising absence of any
correlation between $L_{\rm IR}/L_{\rm HCN}$ and $L_{\rm CO}/L_{\rm HCN}$,
yet a still strong correlation between
$L_{\rm IR}/L_{\rm CO}$ and $L_{\rm HCN}/L_{\rm CO}$ (Fig.~1).
This suggests that
the HCN--IR correlation is more physical than the CO--IR correlation
and that the global SF
efficiency depends on the fraction
of the molecular gas in a dense phase ($L_{\rm HCN}/L_{\rm CO}$).
This is somehow reminiscent of the poor HI--IR correlation vs. 
the better CO--IR correlation discoverd more than two decades ago. 

The direct
consequence of  the linear IR--HCN correlation is that the SF law
in terms of {\it dense} gas has a power law index of 1, which is
different from the widely used  Kennicutt (1998) law of a slope 
of 1.4 for the disk averaged $SFR$ as a function of the total 
gas (HI and ${\rm H_2}$). As we show in next section that this 1.4 
law is not unique, neither valid for normal 
spiral galaxies nor for extreme starbursts/ULIRGs.

\section{The Global SF Law in Dense Gas}

The $SFR$--$M_{\rm H_2}$ (IR--CO) is essentially
linear up to $SFR \sim 20\ms/yr$ (Fig. 2a).  
This seems to also be true in terms of the mean
surface densities of the $SFR$ and molecular  gas mass for the nearest galaxies
with spatially resolved observations (e.g., Wong \& Blitz 2002). 
Thus the linear form of the global SF law in terms of 
total molecular gas density as traced by CO for normal galaxies is 
due to the constant dense gas mass fraction (HCN/CO) in normal 
galaxies (GS04a).  For normal spirals, the SF law is linear 
in terms of both the total 
molecular gas and the dense molecular gas. A fit for 
the normal galaxies in the IR--CO correlation in HCN sample gives 
a slope of 1.0. But there is a poor correlation between the $SFR$ and 
the gas surface 
density in Kennicutt (1998) normal galaxy sample, difficult to derive  
a reasonable slope from normal galaxy sample alone.

 
\begin{wrapfigure}[]{r}{0.55\textwidth}
\vskip -0.2in
\includegraphics[height=2.8truein]{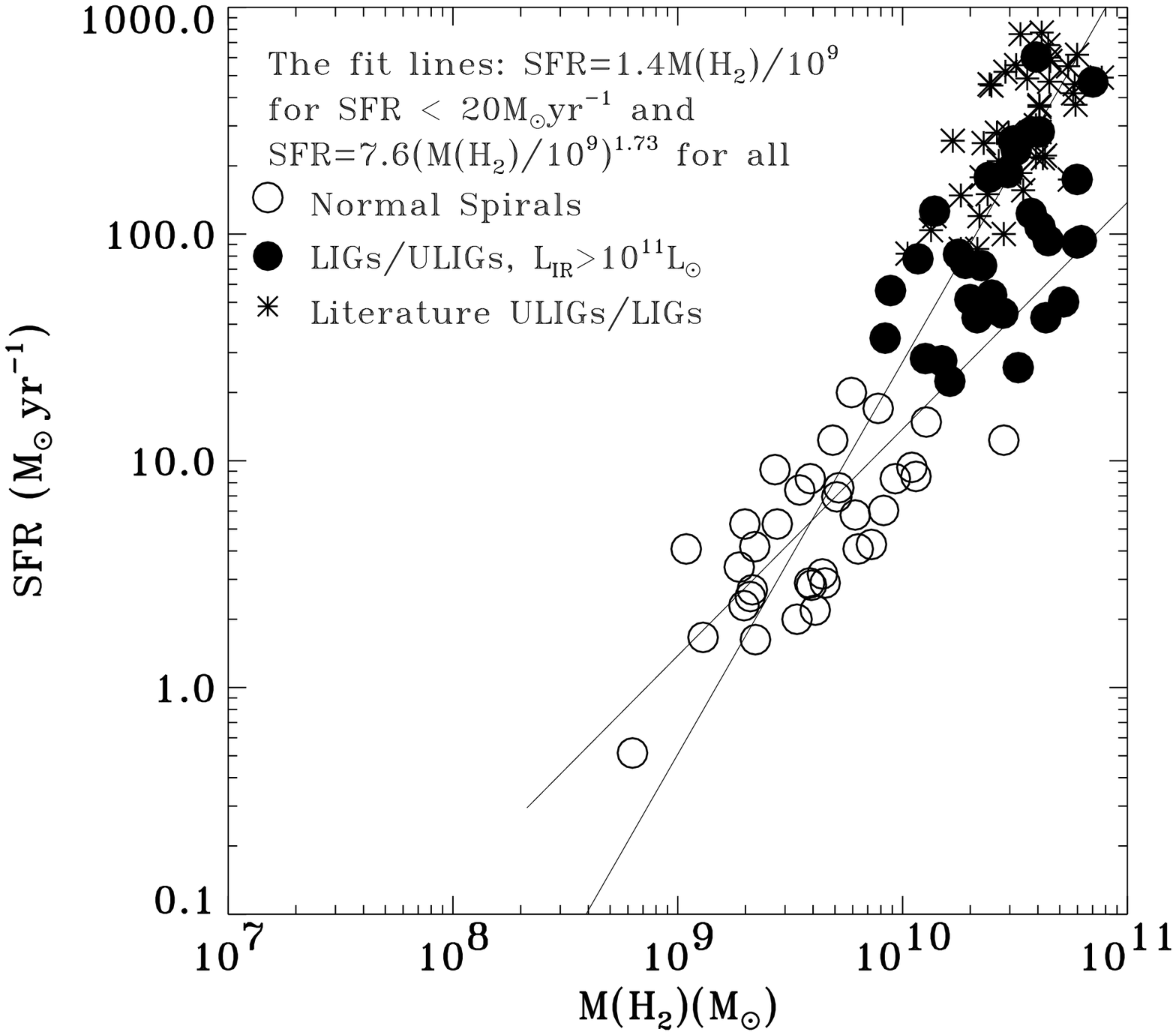}
\includegraphics[height=2.8truein]{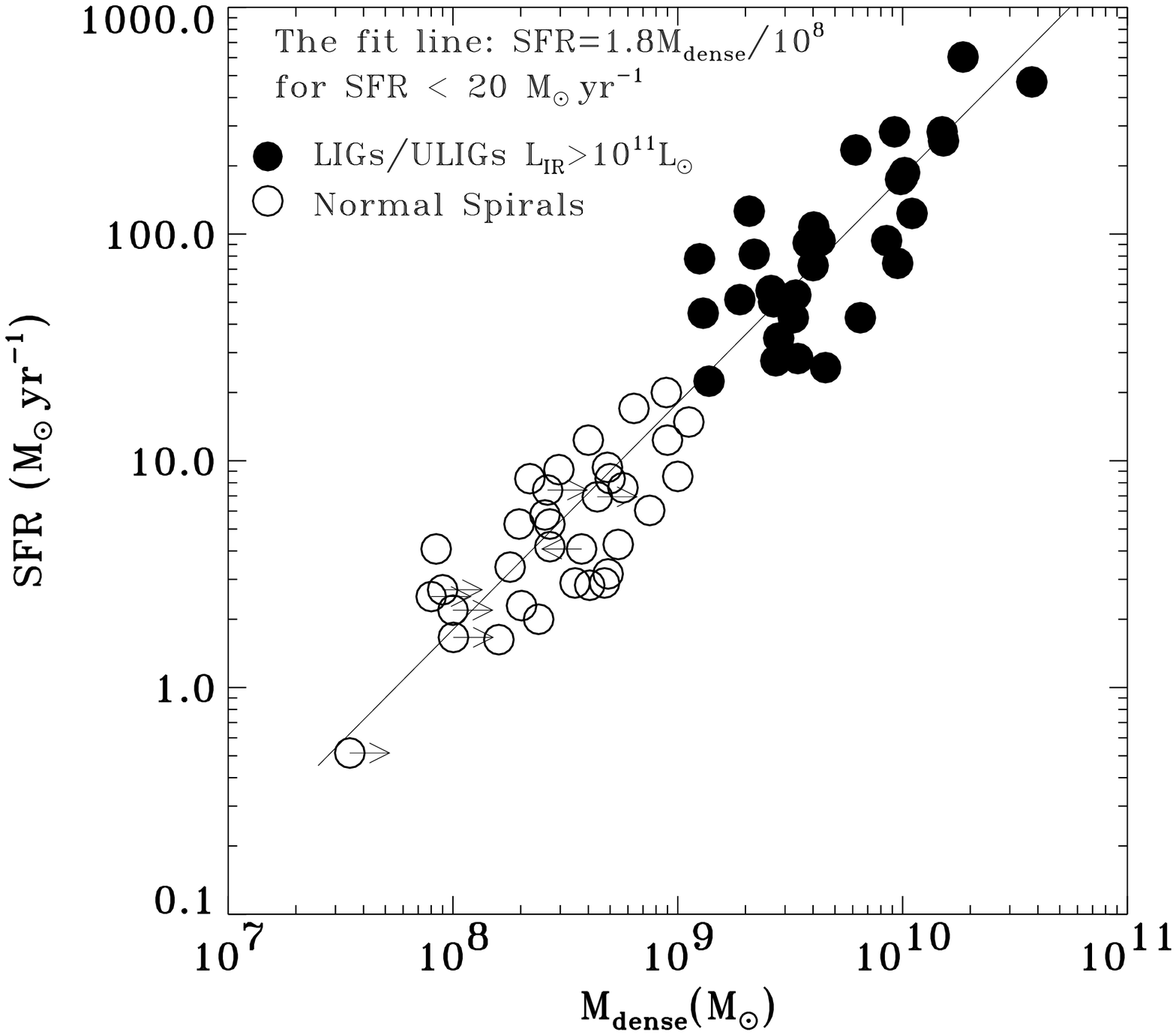}
\vskip -0.1in
\caption{There is no unique SF law in terms of molecular gas or total gas 
(top) since the power index changes with the sample. But we do observe 
a linear SF law in dense gas for all star-forming galaxies.}
\end{wrapfigure}

A direct orthogonal regression fit for all galaxies
in HCN sample leads to a slope of 1.4 (the
least-squares fit slope is 1.3). These fits
are almost identical to the SF law power index in Kennicutt's
(1998) 36 circumnuclear starbursts sample.  It is obvious from 
Fig.~2a that only galaxies with $SFR$ $>20$ \ms/yr (mostly ULIRGs) 
lie above the slope 1.  The combination of normal galaxies and 
ULIRGs leads to a fit of 1.4. 
Therefore, this slope is not a universal slope at all as it changes 
according to the sample selection. The Kennicutt's (1998) 1.4 slope
is  determined mostly 
from the starburst sample. The circumnuclear starbursts have some of 
the characteristics of ULIRGs, e.g., a high 
dense gas fraction  indicated by high HCN/CO ratio.

When we add more ULIRGs into the sample, the
slope becomes steeper, and the fits lead to a 
slope of 1.7. It is
clear that the ULIRGs steepen the slope of the fit.  
There also appears to be
a trend that some normal spirals with the lowest $\Sigma_{\rm SFR}$ and
$\Sigma_{\rm H_2}$ in Kennicutt's sample tend to lie below the 1.4 fit. 
Adding more extreme galaxies, both ULIRGs and  low luminosity
galaxies, tends to steepen the  slope towards 2. Therefore, it is
difficult to derive a unique 1.4 power law based upon the total molecular gas
or the total gas content.

The SF law in terms of dense gas has a unique slope of 1 (Fig. 2b)
since the global $SFR$  is linearly proportional to the mass of the 
dense molecular
gas. Parameterization in terms of observable mean 
surface densities of the dense
molecular gas and the $SFR$ won't change the slope of 1 as both
quantities are simply normalized by the same galaxy disk area. This
linear SF law in dense gas seems valid for all star-forming galaxies.

 
\begin{wrapfigure}[]{r}{0.55\textwidth}
\vskip -0.12in
\includegraphics[height=2.8truein]{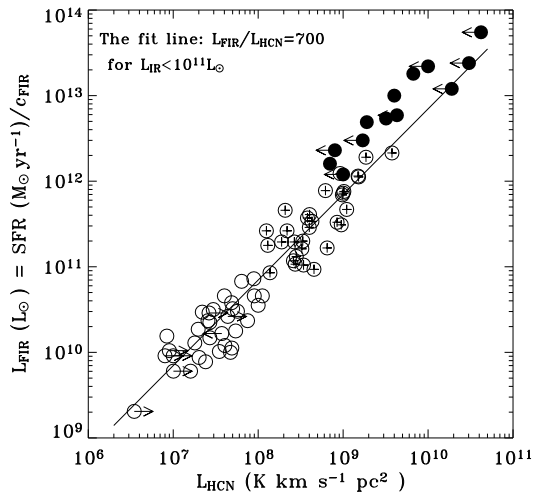}
\vskip -0.1in
\caption{A linear SF law in dense gas appears valid for all star-forming 
systems including galaxies at high-redshit (filled circles). 
The plotted fit is for nearby
normal spirals, and the dense cores of FIR
luminosities of a few times of 10$^{3-6}\ls$  
(not plotted, Wu et al. 2005) lie on the same fit.}
\end{wrapfigure}

Remarkably, the HCN--IR linear correlation is also valid to 
GMC dense cores implying the same physics drives the active SF 
in both dense cores and galaxies (Wu et al. 2005).
New sensitive HCN line observations of four high-redshift
submillimeter (sub-mm) galaxies and QSOs with the VLA including first 
possible HCN detection of a submm galaxy (Gao et al. 2006), 
combined with previous HCN detections and upper limits (e.g., 
Carilli et al. 2005) strongly suggest galaxies at high-redshift
follow this same linear law (Fig. 3). 
The SF law in dense gas appears linear all the way from dense cores
to extreme starbursts at high-z, spanning over nine orders of magnitude
in $SFR$ or FIR luminosity.

The $SFR$ in a galaxy depends linearly on the dense molecular gas
content as traced by HCN, regardless of the galaxy luminosity or the presence
of a ``starburst'', and not the total molecular gas and/or atomic gas
traced by CO and/or HI observations respectively. Since dense molecular 
cloud cores are the  sites of high mass SF, it is the physical 
properties, location and mass of these cores
that set the $SFR$. A detailed SF law can be
determined from observations directly probing the Milky Way cloud 
cores with spatially resolved
measurements and the resolved measurements in nearby star-forming 
galaxies can bridge the dense cores with the galaxies. 
The global $SFR$  of a star-forming system is best characterized by 
its mass of dense gas, $SFR \sim {\rm M(dense~H}_2)$. 
The gas density traced by HCN
emission is apparently near the threshold for rapid SF.

\section{Concluding Remarks}\label{sec:concl}

Although $SFR$ indicated FIR correlates with the various 
cold neutral gas reservoirs (HI, H$_2$, HCN--dense H$_2$),
the FIR-HI is the worst among all and the FIR-CO is non-linear
and has larger scatter than that of FIR-HCN.
The best and tightest FIR-HCN correlation is a linear relation
which implies, under the assumption of the constant conversion factors, 
a linear SF law that $SFR$ is simply proportional 
to the amount of dense molecular gas available to form high mass
stars. Other SF laws based on FIR-CO or FIR-(HI+H$_2$) 
won't have a unique power index. This linear FIR-HCN
relation appears to be valid all the way from dense cores 
in GMCs to extreme starbursts at high-z, revealing the same
physics that massive SF drives most of the energy output
in all these systems.

\begin{acknowledgments}
I wish to express my gratitudes to my collaborators, particularly 
Phil Solomon, for their contributions. China NSF (distinguished 
young scholars) \& Chinese Academy of Sciences
(hundred-elite) are thanked for their supports.
\end{acknowledgments}

\begin{discussion}

\discuss{Kennicutt}{The beautiful aspect of your result is that
the dense cores do not seem to care about whether they reside
in Orion or Arp 220 -- they form stars with the same core efficiency.
But doesn't this simply reword the problems of explaining why the
efficiency of core formation scales with total gas density?}

\discuss{Gao}{No. The dense cores do seem to form massive stars 
with the same efficiency regardless of where they reside. But 
the core formation and the efficiency of core formation 
depend upon the location and/or environment the putative cores
reside. I'm not exactly sure how the
efficiency of core formation scales with total gas density, but 
I'd think the efficiency of core formation scales with the
fraction of gas at high density, rather than total gas or total
molecular gas density. For example, a gas-rich low surface brightness
galaxy could have nearly same disk-averaged total surface gas density 
as that of a starburst galaxy, but the starburst has high fraction 
of molecular gas, particularly dense molecular gas, therefore,
a high star formation rate.}

\discuss{Elmegreen}{I like to interpret these observations in a
different way. It seems the extra power of 1.4 in the Kennicutt
scaling relation is from the rate at which low density gas evolves
toward high density gas. But for observations of only high-density gas,
like yours, their evolution is not needed anymore and then
the power of density is 1, not 1.4. The really important point of
your observations seems to be that there is a universal critical
density for star formation. Your observations are close to this
density.}

\discuss{Gao}{Yes, agree. It seems that a SF law in
terms of low density gas needs extra power of 1.4 or higher. But 
the problem in low density gas (either HI or H$_2$ or both) 
is that it is often difficult to derive a SF law with
a unique power. As I showed that the SF law even in
terms of the total molecular gas does not have a fixed 1.4 power
and the situation is even worse in terms of atomic gas or total gas 
because of the poor correlation between SF rate and 
HI gas. On the other hand, a SF law in terms of total
molecular gas (thus low density) for normal spirals does have
a power index of 1, same as that in terms of dense molecular gas
probed by HCN observations. But the power index in terms of total
molecular gas needs to be increased when extreme (ultraluminous 
and/or low surface brightness) galaxies are added/considered.}

\end{discussion}

\end{document}